\documentclass{iau}
\usepackage{graphicx}

\def\sgr{Sgr\,A$^\star$}

\def\fe{Fe~K$\alpha$}

\title[Reflection of two past outbursts of Sagittarius A$^\star$] 
{The reflection of two past outbursts of Sagittarius A$^\star$ observed by Chandra during the last decade}

\author[Ma{\"i}ca Clavel, R. Terrier, A. Goldwurm et al.]   
{Ma{\"i}ca Clavel$^{1,2}$,
R{\'e}gis Terrier$^1$,
Andrea Goldwurm$^{1,2}$,
Mark R. Morris$^{3}$,
Gabriele Ponti$^{4}$,
Simona Soldi$^{1}$
 \and
Guillaume Trap$^{5,1,2}$
 }

\affiliation{$^1$AstroParticule et Cosmologie, Univ. Paris Diderot, CNRS/IN2P3, CEA/DSM, Observatoire de Paris, Sorbonne Paris Cit{\'e} ; 10, rue A. Domon et L. Duquet, 75205 Paris Cedex 13, France \\ email: {\tt maica.clavel@apc.univ-paris7.fr} \\[\affilskip]
$^2$ Service d'Astrophysique/IRFU/DSM, CEA Saclay, 91191 Gif-sur-Yvette Cedex, France \\
$^3$ Dept. of Physics \& Astronomy, University of California, Los Angeles, CA 90095-1547, USA\\
$^4$ Max-Planck-Institute for Extraterrestrial Physics, PSF 1312, D-85741 Garching, Germany\\
$^5$Unit{\'e} de physique, Palais de la d{\'e}couverte - Universcience, 75008 Paris, France}

\pubyear{2013}
\volume{303}  
\pagerange{344--348}
\jname{The\,Galactic\,center:\,Feeding\,and\,feedback\,in\,a\,normal\,galactic\,nucleus}
\editors{L.\ O.\ Sjouwerman, C. C. Lang \& J. Ott}
\begin{document}

\maketitle

\begin{abstract}
The supermassive black hole at the Galactic center, Sagittarius A$^\star$, has experienced periods of higher activity in the past. The reflection of these past outbursts is observed in the molecular material surrounding the black hole but reconstructing its precise lightcurve is difficult since the distribution of the clouds along the line of sight is poorly constrained.\\
\indent
Using Chandra high-resolution data collected from 1999 to 2011 we studied both the 6.4 keV and the 4--8 keV emission of the region located between \sgr\ and the Radio Arc, characterizing its variations down to 15$''$ angular scale and 1-year time scale. The emission from the molecular clouds in the region varies significantly, showing either a 2-year peaked emission or 10-year linear variations. This is the first time that such fast variations are measured. Based on the cloud parameters, we conclude that these two behaviors are likely due to two distinct past outbursts of \sgr\ during which its luminosity rose to at least 10$^{39}$ erg s$^{-1}$.
\keywords{reflection nebulae, X-rays: ISM, radiation mechanisms: nonthermal, Galaxy: center}
\end{abstract}

\firstsection 
\section{Introduction}
Sagittarius A$^\star$ is the supermassive black hole located in the center of the Galaxy.
Its \hbox{X-ray} luminosity is about $10^{33-34}$~erg~s$^{-1}$ at quiescence
 (\cite{Baganoff2003}) but varies, showing rapid flares (\cite{Neilsen2013}) during which its luminosity remains at least eight orders of magnitude lower than its Eddington luminosity.
Nevertheless, there are strong hints that \sgr\ experienced a large phase of activity in the past (\cite{Ponti2012}) and its recent history can be reconstructed from the non-thermal emission emanating from the molecular clouds at the Galactic center. This emission is characterized by a continuum component and a strong \fe\ line at 6.4~keV.  
The strong variability of this emission, detected in both Sgr~B2 (\cite{Inui2009, Terrier2010, Nobukawa2011}) and the Sgr~A region (\cite{Muno2007, Ponti2010, Capelli2012} et al.\ 2012, \cite{Clavel2013}), proves that a significant fraction of the diffuse emission correlated with the molecular clouds is due to reflection. This reflected emission is created by Compton scattering and K-shell photo-ionization of neutral iron atoms produced by intense \hbox{X-ray} radiation such as emitted during a putative past large flare of \sgr\ (\cite{Sunyaev1993, Koyama1996, Sunyaev1998}). 
Unfortunately, reconstructing the lightcurve of \sgr\ is very complex because the distribution of clouds along the line of sight is barely known. Nonetheless, the temporal behavior of the illumination is directly linked to the clouds' position along the line of sight, so the fine X-ray variations can also be used to constrain the matter distribution and thereby to further characterize the past activity of \sgr. This is why we took advantage of \textit{Chandra}'s high spatial resolution to measure the fine X-ray variations occurring in a key region located between \sgr\ and the Radio Arc, hereafter called the Sgr~A complex. We present here our main results concerning the temporal variations observed in this region and the constraints we derived on the past activity of \sgr. For more details, refer to \cite[Clavel et al.\ (2013)]{Clavel2013}.

\vspace{-0.15cm}
\section{Temporal variations in the Sgr A complex: two time behaviors}\label{sec:2}
\begin{figure}[h]
	\vspace{-0.3cm}
	\centering
	\includegraphics[width=1\textwidth]{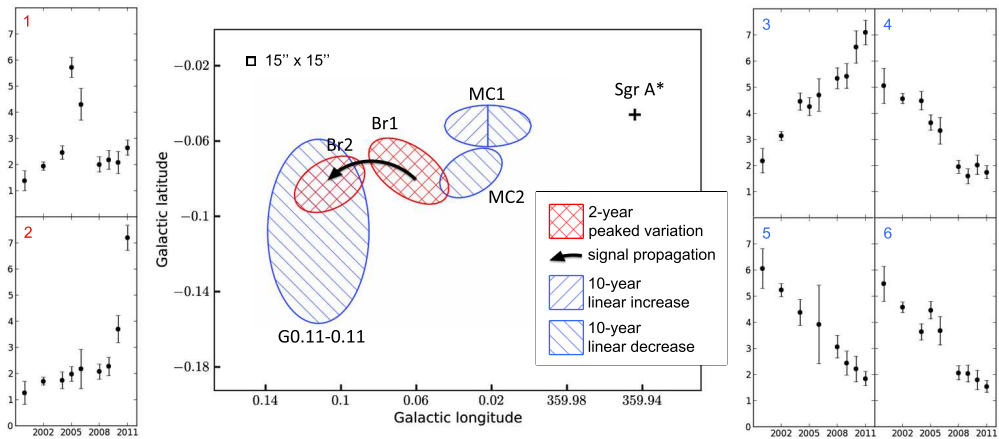}
	\caption{(\textit{Center}) Schematic view of the variations observed in the five main clouds of the Sgr~A complex. (\textit{Left and Right}) Typical 4--8~keV flux lightcurves (lc) in units of 10$^{-6}$~ph~cm$^{-2}$~s$^{-1}$ extracted from 15-arcsec-squared regions of, from 1 to 6, Br1, Br2, MC1-left, MC1-right, G0.11-0.11 and MC2. Linear variations on 10-year time scales are seen in all clouds except for Br1 and Br2, which have been experiencing faster variations.}
	\label{fig:schema}
\end{figure}

Our analysis is based on all the \textit{Chandra} observations covering the ten-arcmin-squared region centred on (l,b)=(0.06$^\circ$, 0.10$^\circ$) that have been collected between 1999 and 2011. The data have been reduced using \textit{Chandra} standard analysis and we used the 4--8~keV flux of 15-arcsec-squared regions in order to characterize the variations of the Sgr~A complex.  This systematic analysis of the region is in perfect agreement with the 6.4~keV characterization also performed for some specific clouds in \cite[Clavel et al.\ (2013)]{Clavel2013} and shows that the five main molecular clouds of the region are varying with a significance of at least $10\,\sigma$ but with two different types of temporal behavior: the variations are either short (year) and peaked, or slower (decade) and linear (Fig.\,\ref{fig:schema}).

{\underline{\it Two-year peaked variations:}}
Br1 and Br2 have mainly been illuminated since 2008, with the illumination starting on the Galactic west of the cloud with a superluminal propagation of the echo away from \sgr\ (\cite{Ponti2010}). Our analysis confirms both the propagation along the Bridge, with the detection of a bright filament at the center of Br2 in 2011 (Fig.\,\ref{fig:schema}, lc~2), and a required \sgr\ luminosity of at least a few times 10$^{39}$ erg s$^{-1}$ in order to account for the illumination. Furthermore, we report the first detection of a flux increase followed by a similar decrease in several subregions of the Br1 cloud (Fig.\,\ref{fig:schema}, lc~1), which forces the duration of the reflection episode to be about two years or less. Since the variation of the reflected emission cannot be faster than the illuminating event itself, this new measurement implies that the past X-ray radiation responsible for the Br1 and Br2 illumination peak lasted no more than two years. 

{\underline{\it Ten-year linear variations:}}
the variations of the three other clouds are characterized by either a ten-year linear increase (Fig.\,\ref{fig:schema}, lc 3) or a ten-year linear decrease (Fig.\,\ref{fig:schema}, lc 4-6). In the following section we discuss whether these slower variations are compatible with the two-year upper limit for the illuminating event derived from the fastest variations. %

\vspace{-0.15cm}
\section{Constraints on \sgr's past activity: two past outbursts}\label{sec:3}
\begin{figure}[h]
	\vspace{-0.3cm}
	\centering
	\includegraphics[trim = 20mm 0mm 20mm 0mm, clip, width=1\textwidth]{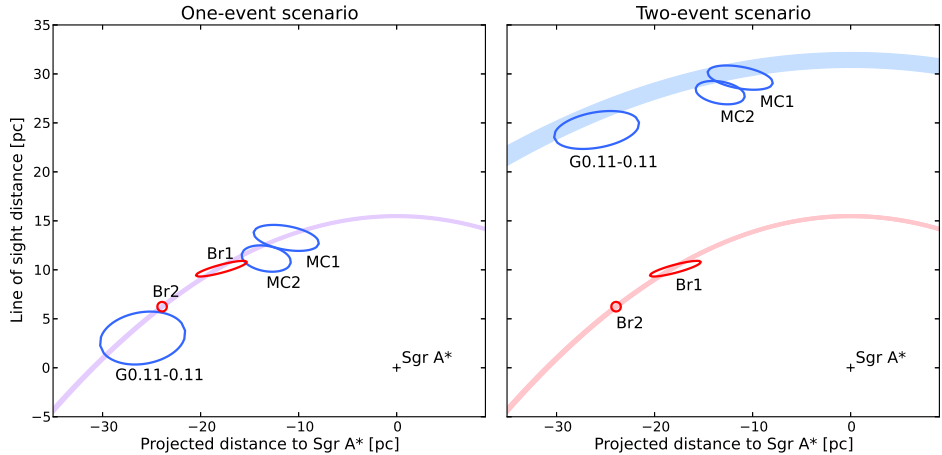}
	\caption{Face-on view of the Galactic center for two different scenarios. In each plot, the ellipses provide possible positions and extensions of the clouds along the line of sight while the parabolas represent the area illuminated by the event(s) at a given date (\cite{Sunyaev1998}). (\textit{Left}) Single-event scenario: a two-year event ending a hundred years ago. As explained in the text, this scenario is excluded. (\textit{Right}) Two-event scenario: a two-year event ending a hundred years ago plus a ten-year event ending two hundred years ago. We point out that we have no constraint on the age of these two events, so their positions along the line of sight are arbitrary and the short event could even be older than the long one. }
	\label{fig:parabola}
\end{figure}

The variations of the reflection component we observed depend on the lightcurve of the illuminating event but also on the structure of the reflecting clouds. So the two different behaviors observed could be due to either two different cloud structures or to two different illuminating flares (Fig.\,\ref{fig:parabola}).

{\underline{\it One-event scenario:}}
to explain the variations seen in the Sgr~A complex, a single-event scenario was first assumed by \cite[Ponti et al.\ (2010)]{Ponti2010}, suggesting an event duration of a few centuries. We now have a strong upper limit of two years on the flare duration, along with more detailed lightcurves of the cloud emissions, which gives stringent constraints for this scenario. First, since the event is so short, all the illuminated clouds ought to lie along the same parabola (Fig.\,\ref{fig:parabola}, \textit{Left}). However, this would imply that a large fraction of the molecular material of the Sgr~A complex is roughly at the same position along the line of sight despite its different molecular line velocities (\cite{Jones2012}). Second, in case of a two-year event, the longer illumination seen in MC1, MC2 and G0.11-0.11 can only be explained by a larger depth of these clouds along the line of sight so that the event parabola moving away from \sgr\ reached them first and illuminates them longer than the Br1 and Br2 clouds. In this case, the similarity of the decreasing trends observed in the three clouds (Fig.\,\ref{fig:schema}, lc 4-6) can only be attributed to similar density distributions within these clouds. Third, knowing the fluence of the reflected emission from each molecular cloud as well as the clouds' relative positions, we can derive an estimation of their column densities. We find that the clouds displaying slower variations should have higher column densities than the Br1 and Br2 clouds. Since this is not what is deduced from the molecular line observations (\cite{Jones2012}), the one-event scenario is excluded.

{\underline{\it Two-event scenario:}}
in addition to the two-year event already described, we have to introduce a second event responsible for the longer variations seen in three of the molecular clouds (Fig~\ref{fig:parabola}, \textit{Right}). Since the slopes of the linear decrease seen in these clouds are very similar we can reasonably assume that this variation pattern is mainly dominated by the time scale of the illuminating event rather than by each cloud's specific configuration. In this case we can set a ten-year lower limit on the duration of the second flare. We have no constraints on the relative position in time of these two events. However, the three clouds experiencing the longer event have similar brightnesses and similar column densities which means that they ought to be about at the same distance from \sgr\ and only old flares (about 200 years old or more) can fulfill this criterion. Since both the position of the cloud along the line of sight and the opacity of the clouds on scales that match the X-ray data are unknown, having an estimate of the events' luminosity is also difficult. Nevertheless, assuming reasonable parameters for the clouds we find that two past flares of \sgr\ with a luminosity of a few 10$^{39}$~erg~s$^{-1}$ and lasting for two and ten years are sufficient to explain the short and long behaviors, respectively.

\vspace{-0.15cm}
\section{Conclusions}
Our systematic analysis of the flux variations of the Sgr A complex, divided into 15-arcsec-squared subregions, reveals significant variations in all the molecular clouds of the complex. We report two different types of behavior: on the one hand a two-year peaked variation which is the fastest variation detected so far in the diffuse X-rays from Galactic center molecular clouds, and on the other hand a ten-year linear variation. Based on these results we are able to exclude a single-event scenario and we can provide constraints on the luminosity ($>$ few 10$^{39}$~erg~s$^{-1}$) and duration ($<2$ and $>10$ years, respectively) of the two past flares of \sgr. These new results will provide key evidence for identifying the physical processes responsible for the past changes in the luminosity of \sgr.

\end{document}